\documentclass{aastex}
\usepackage{spr-astr-addons}
\usepackage{url}

\RequirePackage{color}

\begin{document}

\title{A study of the scaling relation $M_{\bullet }\propto R_{e}\sigma ^{3}$ for supermassive black holes and an update of the corresponding theoretical model}
\shorttitle{``A study of the scaling relation $M_{\bullet }\propto R_{e}\sigma ^{3}$"}
\shortauthors{Beltramonte, Benedetto, Feoli, Iannella}

\author{T. Beltramonte\altaffilmark{1}}
\email{tbeltram@unisannio.it}
\and
\author{E. Benedetto\altaffilmark{1,2}}
\email{elmobenedetto@libero.it}
\and
\author{A. Feoli\altaffilmark{1}}
\email{feoli@unisannio.it}
\and
\author{A.L. Iannella\altaffilmark{1}}
\email{antonellalucia.iannella@unisannio.it}

\altaffiltext{1}{Department of Engineering, University of Sannio, Piazza Roma 21, 82100 Benevento, Italy}
\altaffiltext{2}{Department of Computer Science, University of Salerno, Via Giovanni Paolo II, 132, 84084 Fisciano (Sa), Italy}

\begin{abstract}
In this paper we want to compare the theoretical predictions of a law
    proposed by Feoli and Mancini, with the most recent experimental data about
    galaxies and Supermassive black holes. The physical principle behind this law is the transformation of the angular
    momentum of the interstellar material, which falls into the black hole, into
    the angular momentum of the radiation emitted in this process. Despite the
    simplicity of the model, this law shows an excellent agreement with the experimental
    data for early - type galaxies while a new approach is proposed for spirals.
\end{abstract}

   \keywords{supermassive black holes; early-type galaxies;
spiral galaxies; radiative efficiency}


%

\section{Introduction}

Thanks to the very high angular resolution of modern telescopes, it is
increasingly evident that nearby galaxies ($<100$ Mpc) have a supermassive
black hole (SMBHs; $M_{\bullet }>10^{6}M_{\odot}$) at their center \citep{Ferrarese2005,Kormendy1995,Richstone1998}.
Equally evident is the existence of a strong correlation between the mass
of these SMBHs and the properties of the galaxies that host them \citep{Aller2007,Beifiori2009,Feoli2005}, \citep{Feoli2009,Graham2007,Gultekin2009a,Haring2004,Hu2008,Kisaka2008,Magorrian1998,Mancini2012,Marconi2003,Tremaine2002}. For
this reason, it is clear that there is a link between the process of
accretion of SMBHs and the formation and evolution of their galaxies \citep{Begelman2005,Burkert2001,Cavaliere2002,Croton2006,Granato2004,Haiman2004,Hopkins2007a,King2003,Murray2005,Pipino2009a,Pipino2009b,Sazonov2005,Silk1998,Sijacki2007,Wyithe2003}.
Furthermore, numerical simulations and semi-analytical models seem to reveal
that the environment, in which these galaxies live, can greatly influence
the growth and activity of SMBHs \citep{Booth2009,Ciotti2009,Degraf2011,DiMatteo2005,Fanidakis2011,Feoli2011b,Hopkins2008,Shin2010,Shin2012}.

Over the past twenty years, the astrophysicists have found a large number of scaling
laws, in which the mass of the central SMBH correlates with different properties of the
host spheroidal component, such as the bulge luminosity, mass, effective
radius, central potential, dynamical mass, concentration, Sersic index,
binding energy, kinetic energy of random motions, X-ray luminosity, momentum
parameter, etc. \citep{Ferrarese2009,Feoli2007,Gebhardt2000,Graham2005,Gultekin2009b,Hopkins2007b,Laor2001,Merritt2001,Soker2011,Wandel2002}.
Recently,  a possible relation between $%
M_{\bullet }$ and the number of globular clusters has also been hypothesized \citep{Burkert2010,Snyder2011}. Instead, with
regard to the correlation with the halo of dark matter, there is still a
wide debate \citep{Baes2003,Bellovary2011,Ferrarese2002,Feoli2011a,Kormendy2011,Volonteri2011}.

In this paper, starting from the conservation of the angular
momentum, we want to analyze another scaling law that was inspired by the
numerical results of \citep{Hopkins2007a}, but it is based on a theoretical framework
\citep{Feoli2011a} : the $M_{\bullet }-R_{\mathrm{e}}\sigma ^{3}$ law where $R_{%
\mathrm{e}}$ and $\sigma $ are the effective radius and the velocity
dispersion of the bulge, respectively. This law seems to have an excellent
agreement with experimental data and it could be a basic law for galaxies
and their SMBHs.

\section{The Sample}

One of the largest database of Black Hole masses is certainly the one
created by van den Bosch \citep{vandenBosch2016} who has collected data for 230 galaxies.
In order to obtain a fit with a reliable accuracy, we have preferred to
exclude the masses of Black Holes measured with a relative error on $\log
(M_{\bullet })$ greater than or equal to 1. In this way, we obtained a sample
of 181 galaxies. The galaxies that have been neglected are listed in Appendix A.
 We have also
removed from the sample NGC4486b, because this galaxy ``deviates strongly from
every correlation involving its black hole mass" \citep{Saglia2016}, and NGC404
because its mass is too low for a supermassive black hole $ M_{\bullet} < 10^{6}M_{\odot}$.
We have studied the remaining sample of 179 objects with a particular attention for the
behavior of spiral galaxies in the context of our relation. In general, they
have a rotation velocity which is greater than the velocity dispersion (but NGC 4388 is a typical counterexample having $V/\sigma = 0.6$)
and our theoretical model works well for low values of angular momentum, so
we have verified in a first approach \citep{Feoli2019} that the fit  of the experimental data
 improves if we exclude from the sample all the spiral galaxies. In order to face the problem, in this paper the
sample has been divided into two subsamples,  one with all the 66 spiral galaxies listed in
Appendix A and the other including the remaining  113 early-type galaxies that have been studied separately.
 Our aim is to motivate the experimental necessity to
update  our theoretical model for spirals and  to propose the suitable corrections.

\section{The Theoretical Model}

Let us remember that the core of Feoli-Mancini's model \citep{Feoli2011a} is the
transformation of the angular momentum of the matter  falling into the
black hole, into the angular momentum of the radiation emitted during the
process. In this scenario we have a flux of gas with an internal speed $%
V_{in}=\sigma $ and a supermassive black hole of mass $M_{\bullet }$ that
grows emitting jets of radiation. Therefore, to describe the system, we need
both a suitable velocity field of the gas falling into the black hole and
the conservation of the angular momentum. We start by stating that the
velocity of the gas $V_{part}$ in the galactic bulge is linked to the
stellar velocity dispersion $\sigma $ in such a way that the radial
component is $V_{in}=\sigma $ \citep{Soker2011,Soker2009} and the transverse
component is the  rotational velocity $V_{rot}$ of the
gas. Furthermore, if we assume that each emitted photon has a velocity $c$
in the same direction as the incoming gas particle $V_{part}$ and it forms
an angle $\alpha $ with the accretion line and with the radial component of
gas velocity $V_{in}$ (for details and for a useful figure you can read \citet{Feoli2014}), we get $V_{rot}/V_{in}=\tan(\alpha) $. Looking for an estimate of the
angular momentum for the galactic bulge, we obtain the conservation
equation:
\begin{equation}
M_{acc}R_{e}V_{rot}=M_{acc}R_{e}V_{in}\tan(\alpha) =cM_{rad}R_{A}\sin(\alpha)
\end{equation}%
where $R_{A}$ is the accretion radius of the SMBH, which can be estimated
through the Bondi-Hoyle-Lyttleton (BHL) theory \citep{Bondi1944,Hoyle1939}, that
predicts a value
\begin{equation}
R_{A}=2GM_{\bullet }/V_{in}^{2}
\end{equation}%
where $G$ is the gravitational constant. It is interesting to note that in
this case with $V_{in}=\sigma $, the BHL radius coincides with the so called
radius of influence of the Black Hole.\\
\newline
{\bf a) Case $\sigma >> V_{rot}$}

 As Bondi-Hoyle-Lyttleton theory (BHL)
is a realistic model in the case of radial growth with low transverse
velocity, it is possible to consider $\tan(\alpha) \simeq 0$ getting $\tan(\alpha)
\simeq \sin(\alpha)$. Finally, substituting (2) in (1), the mass of the black
hole $M_{\bullet }$ can be written in this form:
\begin{equation}
M_{\bullet }=\frac{R_{e}\sigma ^{3}}{2\epsilon cG}
\end{equation}%
where
\begin{equation}
\epsilon =M_{rad}/M_{acc}
\end{equation}%
is the captured mass to radiative energy conversion efficiency.
By considering the logarithm of the equation (3), we get
\begin{equation}
\log\left( \frac{M_{\bullet }}{M_{\odot}}\right) =\log\left( \frac{%
1}{2\epsilon }\right) +\log\left( \frac{R_{\mathrm{e}}\,\sigma ^{3}}{%
c\,G\,M_{\odot}}\right)
\end{equation}%
In order to test our theoretical model, we need to get from the experimental data the corresponding best fit line
\begin{equation}
y=b+mx
\end{equation}%
where
\begin{equation}
y=\log\left( \frac{M_{\bullet }}{M_{\odot}}\right)
\end{equation}%
and
\begin{equation}
x=\log\left(\frac{R_{\mathrm{e}}\,\sigma ^{3}}{c\,G\,M_{\odot}}\right)
\end{equation}%
Finally, we have to compare the parameters ($m$ and $b$) with the theoretical
predictions that are $m=1$ and
\begin{equation}
b=\log\left( \frac{1}{2\epsilon }\right).
\end{equation}%
However,  this approach, contained in previous papers, is valid only for early-type galaxies that have a low rotation velocity, and it will be evident
from the statistical elaboration of data. Therefore, we propose a corrected version of the model that we expect to be valid for spirals.\\
\newline

{\bf b) Case $\sigma << V_{rot}$}

 We suppose that $V_{part}$ has practically the same direction of $V_{rot}$ hence we can consider now
\begin{equation}
\left\{
\begin{array}{c}
\overrightarrow{R}_{A}\bot \overrightarrow{V}_{rot} \\
\overrightarrow{R}_{A}\bot \overrightarrow{c} \\
\overrightarrow{R}_{e}\bot \overrightarrow{V}_{rot}%
\end{array}%
\right.
\end{equation}
and the conservation of angular momentum becomes
\begin{equation}
M_{acc}V_{rot}R_{e}=cM_{rad}R_{A}
\end{equation}
For the rotation velocity, we can choose the newtonian one
\begin{equation}
V_{rot}=\sqrt{\frac{GM_{G}}{R_{e}}}
\end{equation}
(where $M_G$ is the bulge mass), and for the accretion radius, the black hole radius of influence
\begin{equation}
R_{A}=2GM_{\bullet }/\sigma ^{2}.
\end{equation}
Substituting (13) and (12) in (11) we obtain
\begin{equation}
\sqrt{\frac{GM_{G}}{R_{e}}}R_{e}=\epsilon c\frac{2GM_{\bullet }}{\sigma ^{2}}
\end{equation}
By dividing and multiplying by $\sqrt{M_{\bullet }}$ and squaring, we have
\begin{equation}
M_{\bullet }=\left( \frac{M_{G}}{M_{\bullet }}\frac{\sigma }{c}\right) \frac{%
1}{4 \epsilon ^{2}}\left( \frac{R_{e}\sigma ^{3}}{cG}\right)
\end{equation}
and
$$\log (M_{\bullet })=-\left[ \log (M_{\bullet })-\log \left(\frac{M_{G}\sigma }{c}\right)%
\right] -\log \left( 4\epsilon ^{2}\right)+$$
\begin{equation}
+\log \left( \frac{R_{e}\sigma
^{3}}{cG}\right)
\end{equation}
But Soker and Meiron \citet{Soker2011} have found a relation
\begin{equation}
\log (M_{\bullet })= d + p \log \left(\frac{M_{G}\sigma }{c}\right)
\end{equation}
hence, in the simplest case of $p \simeq 1$, we can write the equation (16) in the form
\begin{equation}
\log (M_{\bullet })=-d-2\log \left( 2\epsilon \right) +\log \left( \frac{%
R_{e}\sigma ^{3}}{cG}\right)
\end{equation}
and we have to study for spiral galaxies a relation
\begin{equation}
\log (M_{\bullet })=\beta +\alpha \log \left( \frac{R_{e}\sigma ^{3}}{cG}%
\right)
\end{equation}
expecting for the parameters the values
 $\alpha =1$ and
\begin{equation}
\beta = -d -2\log \left( 2\epsilon \right)
\end{equation}

\begin{table*}[!t]
\footnotesize
\caption{Fit Results of three different Subsamples of Galaxies.}
\begin{center}
{\normalsize
\begin{tabular}{|c|c|c|c|c|c|c|}
\hline
& \multicolumn{3}{|c|}{\textbf{LINMIX\_ERR}} & \multicolumn{3}{|c|}{\textbf{MPFITEXY}} \\ \cline{2-7}
\textbf{$y=b+mx$} & \textit{Subsample} & \textit{Subsample} & \textit{Subsample} & \textit{Subsample} & \textit{Subsample} & \textit{Subsample} \\
& \textit{of 179} & \textit{of 113} & \textit{of 66} & \textit{of 179} & \textit{of 113 } & \textit{of 66}\\ \hline
$m$ & $1.12\pm 0.07$ & $1.00\pm 0.07$ & $1.23\pm 0.21$ & $1.13\pm 0.07$ & $1.01\pm 0.06$ & $1.23\pm 0.17$ \\
\hline
$b$ & $-0.045\pm 0.53$ & $0.98\pm 0.51$ & $-1.21\pm 1.47$ & $-0.15\pm 0.50$ & $0.93\pm 0.47$ & $-1.20\pm 1.22$ \\
\hline
$\varepsilon_0$ & $0.59\pm 0.04$ & $0.47\pm 0.04$ & $0.62\pm 0.07$ & 0.57 & 0.45 & 0.59 \\ \hline
$R$ & 0.765 & 0.825 & 0.640 & -- & -- & -- \\ \hline\hline
\multicolumn{6}{l}{m and b are
the slope and the intercept of the linear relation respectively,}\\
\multicolumn{6}{l}{$\varepsilon_0$ is
the intrinsic scatter of the relationship and $R$ the linear correlation coefficient}%
\end{tabular}
}
\end{center}
\par
{\normalsize \hrulefill \vspace*{4pt} }
\end{table*}

\section{Fit Results}
In order to fit the equation (5) with the data sets, we use three methods: the first is FITEXY, a linear regression routine \citep{Press1992}, that minimizes the $\chi ^{2}$ %


\begin{equation}
\chi ^{2}=\sum_{i=1}^{N}\frac{(y_{i}-b-mx_{i})^{2}}{(\Delta
y_{i})^{2}+m^{2}(\Delta x_{i})^{2}}.
\end{equation}%
To obtain the most efficient and unbiased estimate of the slope, it is necessary to consider
the residual variance in the fitting method, also known as
intrinsic scatter $\varepsilon _{0}$, which is that part of the variance
that cannot be attributed to specific causes \citep{Novak2006}. When the reduced $%
\chi _{\mathrm{r}}^{2}=\chi ^{2}/(N-2)$ of the fit is not equal to 1, we can use MPFITEXY,
an evolution of the routine FITEXY \citep{Tremaine2002}, that automatically normalizes $\chi _{\mathrm{r}}^{2}$
including $\varepsilon _{0}$ in the equation (21):

\begin{equation}
\chi^2_{\mathrm{r}} = \frac{1}{N-2} \sum_{i=1}^{N} \frac{(y_i -b-m x_i)^2}{%
(\Delta y_i)^2 + \varepsilon_{0}^2+ m^2 (\Delta x_i)^2}.
\end{equation}

The second fitting method is the Bayesian linear regression routine LINMIX\_ERR \citep{Kelly2007}
that allows to determine the slope, the normalization, and the intrinsic scatter of the relation
\begin{equation}
\log (M_{\bullet })=b +m\log (x)+\varepsilon _{0}.
\end{equation}%
This routine approximates the distribution of the independent variable as a
mixture of Gaussians, bypassing the assumption of a uniform prior
distribution on the independent variable, which is used in the derivation of
$\chi ^{2}$--FITEXY minimization routine. Since a direct computation of the
posterior distribution is too computationally intensive, random draws from
the posterior distribution are obtained using a Markov Chain Monte Carlo
method \citep{Kelly2007}.
We have analyzed the three samples of 179, 113 and 66 galaxies respectively, and
we have obtained with MPFITEXY and with LINMIX\_ERR the values shown in table 1.
We have also included in the table the Pearson linear correlation coefficient calculated with the formula
\begin{equation}
R=\frac{\sum_{i=1}^{n}(x_{i}-\bar{x})(y_{i}-\bar{y})}{\sqrt{%
\sum_{i=1}^{n}(x_{i}-\bar{x})^{2}}\sqrt{\sum_{i=1}^{n}(y_{i}-\bar{y})^{2}}}%
\;.
\medskip
\end{equation}%

Finally, a third method was used to construct Fig. 1, in which the 179 experimental points
are shown; the colors distinguish spiral galaxies from early-type
galaxies, and the two straight lines indicate the best fit obtained imposing a slope equal to one for the two subsamples fitted separately. To calculate the $b$ intercepts of the subsamples of 113 and 66 objects, we have used the exact formulas of the least squares method \citep{Feoli2007}, in the case $m=1$:
\begin{equation}
b =\frac{\sum_{i=1}^N \bigl(\frac {y_i-x_i}{(\triangle x_i)^{2}+(\triangle y_i)^{2}}\bigr )}{\sum_{i=1}^N \bigl(\frac {1}{(\triangle x_i)^{2}+(\triangle y_i)^{2}}\bigr )}
\end{equation}

\begin{equation}
(\triangle b)^2 =\frac{1}{\sum_{i=1}^N \bigl(\frac {1}{(\triangle x_i)^{2}+(\triangle y_i)^{2}}\bigr )}
\end{equation}
The results are:

\begin{eqnarray}
\nonumber early-type \quad galaxies &\rightarrow& b=1.10 \pm 0.02; \\
\nonumber spiral \quad galaxies &\rightarrow& b=0.55 \pm 0.03.
\end{eqnarray}

\begin{figure*}[!t]
\centerline{\includegraphics{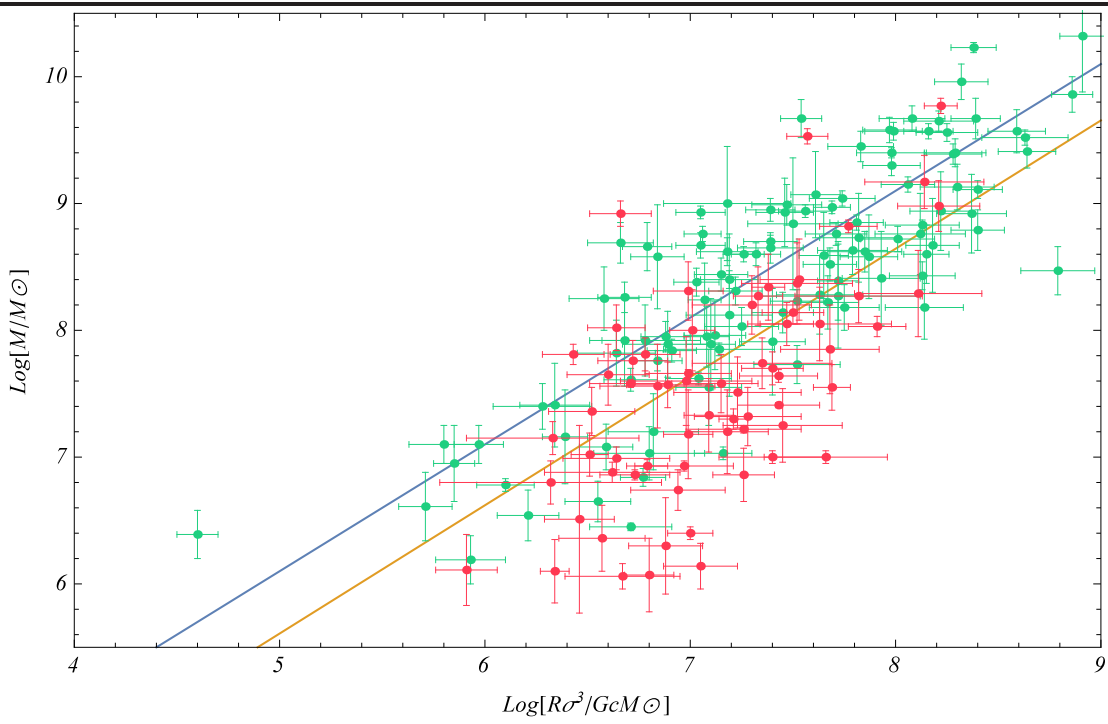}} 
\caption{\itshape {The whole sample of 179 objects in the $\log(M_{\bullet}/M_{\odot})$ vs. $ \log(R_{\mathrm{e}}\,\sigma^{3}/c\,G\,M_{\odot})$ plane.
Red points are spiral galaxies, green points are early-type galaxies and, imposing a slope $m = 1$, the two best fit lines for spiral and early - type galaxies have an intercept $b = 0.55$ (yellow line) and $b = 1.10$ (blue line) respectively.}}
\end{figure*}

 The gap between the two sets of galaxies is evident, hence the necessity to use for them two different theoretical equations.
Furthermore, the linear correlation coefficient and the intrinsic scatter
(Table 1) indicate a stronger relation without spiral galaxies, so the model
works better with early-type galaxies as we expected. For the sample of 179 objects and the sample of 113, the
slope of the best fit line is about 1, as predicted by the theory, and the
value is stable enough to be considered a good result for our test of the
theoretical model. For the subsample of only spirals, the slope  reaches the value of 1.23, starting a little ``tilt" with respect to the unity,
probably due to a slope $p \neq 1$ in the relation (17). Actually, if  $p \neq 1$, the slope of the $M_{\bullet }\propto R_{e}\sigma ^{3}$ relation
from (15) becomes $p(2p-1)^{-1}$. Another possible explanation is the estimation of errors, because the least-squares fit, done using the linear regression routine of Mathematica, without taking into account the errors on the single measurements, gives just $m = 1.05 \pm 0.16$.

Finally,  the intercept parameter cannot be
considered very stable, because it changes with small variations of the
subsample considered for the fit, even if the value found for the 113 early
type galaxies is in the range predicted by the theory.

\begin{figure*}[!t]
\centerline{\includegraphics{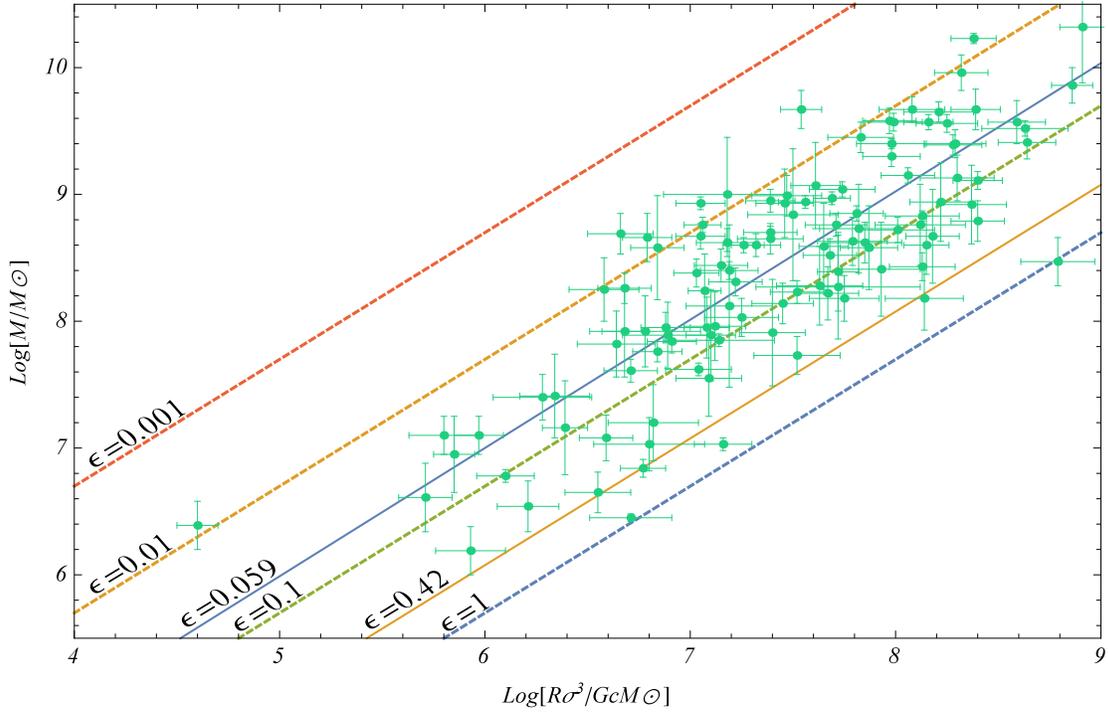}}
\caption{\itshape{The subsample of 113 early - type galaxies inserted in a diagram with lines of different values of the efficiency, according with the equations of the theoretical model. The line with $\epsilon = 0.059$ is the best fit line obtained with MPFITEXY, while the line with $\epsilon = 0.42$ represents the maximum of radiative efficiency for a Kerr rotating Black Hole.}}
\end{figure*}

\section{Predictions}

There are cases in which predicting the mass of a black hole is arduous, sometimes there seems to be no agreement between theoretical models and experimental data, as in the case of NGC1600. We considered this particular case as reported in \citet{Thomas2016}, because they declare: ``The $M_{\bullet}$ in NGC1600 is ... three to nine times more than the percentage predicted from the known scaling relation of black hole and galaxy bulge mass" and  ``the black hole in NGC1600 is ten times more massive than would be expected given the galaxy's velocity dispersion" (Extended Data Figure 1). Hence, according to \citet{Thomas2016}, the value for NGC1600
\begin{equation}
\log (M_{\bullet })=10.23\pm 0.04
\end{equation}
cannot be predicted from the $M_{\bullet}-M_{bulge}$ and $M_{\bullet}-\sigma$ relations, but, in principle, it could be predicted by  our relation. In Appendix B we have explicitly calculated the black hole mass using our relation
obtaining:

\begin{equation}
\log (M_{\bullet})=9.41\pm 0.84,
\end{equation}
and using the $M_{\bullet}-\sigma$ estimated with the same sample obtaining
\begin{equation}
\log (M_{\bullet})=9.26\pm 0.92,
\end{equation}
It is clear that our report foresees the experimental data in the range of errors, unlike \citet{Thomas2016}, so the relation can be a useful tool to predict
black hole masses in difficult cases.

Furthermore, in \citet{Feoli2014},  a classification of black holes based on
the value of the parameter $\epsilon$ has been proposed. This coefficient describes the
efficiency of the black hole in the conversion of the matter captured into
emitted radiation. From (3) we get
\begin{equation}
\epsilon =\frac{R_{e}\sigma ^{3}}{2M_{\bullet }cG}
\end{equation}
and we can classify supermassive black holes following this scheme:

Type $1$ SMBH when $0.1<\epsilon<1.0$ (high efficiency),

Type $2$ SMBH when $0.01<\epsilon<0.1$ (normal efficiency),

Type $3$ SMBH when $0.001<\epsilon<0.01$ (low efficiency).

 Thanks to the efficiency lines of Fig. 2,  we have divided \citet{Feoli2019} the 113, experimental points of early-type galaxies  in this way:
 $30$ (26.5\%) belong to the first type, $75$ (66.4\%) to the second and $8$ (7.1\%) to the third one, even if the error bars on the single variables make  a classification for the galaxies placed on the border line between two different zones of the diagram difficult.

From the fit of the experimental data for early - type galaxies
 and from the prediction of the equation (9), the average value expected for
the efficiency is $\epsilon =0.059$ according to MPFITEXY and $\epsilon
=0.052$ according to LINMIX\_ERR, which are in agreement with the
theoretical models. Our parameter $\epsilon $ is in fact related to the
radiative efficiency $\eta $ defined as $\eta =L/\dot{M}_{\mathrm{acc}}c^{2}$
where $L=\dot{M}_{\mathrm{rad}}c^{2}$ is the radiated luminosity. Actually,
deriving the equation (4) with respect to time,
\begin{equation}
\dot{M}_{\mathrm{rad}}=\dot{\epsilon}M_{\mathrm{acc}}+\epsilon \dot{M}_{%
\mathrm{acc}}
\end{equation}%
and dividing for $\dot{M}_{\mathrm{acc}}$, we have
\begin{equation}
\frac{\dot{M}_{\mathrm{rad}}}{\dot{M}_{\mathrm{acc}}}=\eta =\dot{\epsilon}%
\frac{M_{\mathrm{acc}}}{\dot{M}_{\mathrm{acc}}}+\epsilon
\end{equation}%
If $\epsilon $ can be considered constant during the interval of time of our
astrophysics observations (very small with respect to the black hole life),
its derivative vanishes and $\epsilon =\eta $. From the value of the last stable orbit one can derive
the maximum of radiative
efficiency. For a static Schwarzschild Black Hole  $\eta =0.057$ was found, while
for a Kerr rotating Black Hole the limit is $\eta =0.42$ \citep{Narayan2005}. As the most
part of supermassive black holes rotates, we have drawn in fig. 2 the corresponding line
and it is evident that only three galaxies overcome this
border J0437+2456, NGC2960 (probable spirals), NGC4371, but they are inside the upper limit $%
\epsilon =1$, while one galaxy is just on the last border, the Seyfert
UGC1841, for which probably the model does not work.
\begin{figure*}[!t]
\centerline{\includegraphics{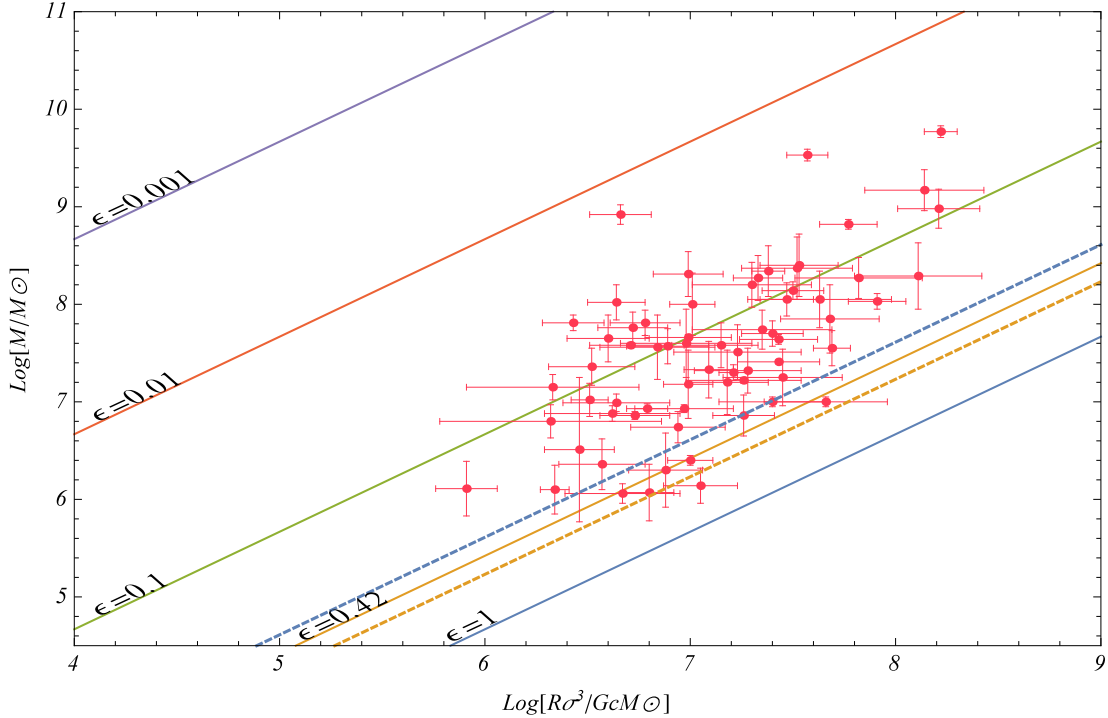}} 
\caption{\itshape {The subsample of 66 objects inserted in a diagram with lines of different values of the efficiency, according with the equations of the theoretical model for spirals. The border for Kerr Black holes $\epsilon = 0.42$ is also shown using $d = 0.73$ together with the two dashed lines given by the error on $d$ of $\pm 0.19$ (see equations (17), (20) and (33) in the text).}}
\end{figure*}

An analogue classification for spiral galaxies is more complicated. It is necessary either to know the bulge mass of each galaxy and to use directly the
equation (15), or to choose a suitable best fit of the relation (17),  for example the one published in \citet{Feoli2011a}
\begin{equation}
\log (M_{\bullet })=
 (0.73 \pm 0.19)+ (1.01 \pm 0.03) \log \left(\frac{M_{G}\sigma }{c} \right)
\end{equation}
from which to extract the value of the intercept $d$ that appears in the equation (20). Of course this second way is affected by a greater error in the estimate of $\epsilon$, but it allows to draw the four lines of Fig.2 for $\epsilon = 1, 0.1, 0.01, 0.001$ also on the corresponding figure for spiral galaxies (Fig.3). Using the relation (33) we can divide the black holes of spiral galaxies into three types counting $43$ objects (65\%) of the first type, $23$ (35\%) of the second and zero of the third one. It means that on average the supermassive black holes of spirals are more efficient with respect to the ones of early - type galaxies. Even if all the 66 points in fig. 3 are well inside the line of $\epsilon = 1$, we have analyzed the behavior near the $\epsilon = 0.42$ line that has become a border zone between two dashed lines (Fig. 3) that represents the uncertainty on the intercept $\pm 0.19$ in the relation (33). The result is that only one object is outside (NGC 4945) and other five galaxies are inside the border zone: Circinus, NGC1386, NGC3079, NGC5248 and NGC5495.

\section{Conclusions}
Using the data collected by van den Bosch \citep{vandenBosch2016} we have tested the $M_{\bullet }\propto R_{e}\sigma ^{3}$ scaling relation. Our fit shows that surely this correlation exists, even if it is weak for the sample of only spirals. Anyway the correlation can be used to predict the black hole mass in difficult cases, such as NGC1600.

 Furthermore, we have found that our old very simple theoretical model \citep{Feoli2011a} is consistent with the sample of data for early - type galaxies both for the slope and the intercept of the best fit line and can be used to derive useful predictions about the efficiency of supermassive black holes. On the other side, it is necessary an update for the case of a velocity rotation greater than the dispersion velocity. We have proposed a new theoretical approach and we have checked whether it works with a subsample of 66 spiral galaxies. The results are encouraging but not definitive. On  average, the efficiency of Black Holes in spirals is greater than the one of early - type galaxies, but also the number of galaxies on the limit of maximum efficiency, with the risk that the model does not work is not negligible (6 over 66 objects). It means that the model for spirals must be further  verified with another larger sample.

\section*{\footnotesize Acknowledgements}
{\footnotesize The authors thank Franco Caprio for his help with the software and Remco C. E. van den Bosch for his database.\\
This research was partially supported by FAR fund of the University of Sannio.}

\appendix
\section{List of galaxies}

 The galaxies discarded because the black hole mass is known with a relative error on $\log (M_{\bullet}) \geq 1$ are the following
 \newline
\newline
NGC0205, NGC0289, NGC0428, NGC0598, NGC1042, NGC1428, NGC1493, NGC2685,
NGC2778, NGC2903, NGC2964, NGC3021, NGC3310, NGC3351, NGC3423, NGC3621,
NGC3642, NGC3675, NGC3945, NGC3982, NGC4088, NGC4150, NGC4212, NGC4245,
NGC4314, NGC4321, NGC4382, NGC4435, NGC4450, NGC4579, NGC4800, NGC5347,
NGC5427, NGC5457, NGC5643, NGC5879, NGC5921, NGC6300, NGC6503, NGC6951,
NGC7418, NGC7424, UGC1214, UGC1395, UGC9799, Henize2-10, IC0342, IC3639,
Mrk1216. \newline
\newline
The spiral galaxies enclosed in the subsample of 66 objects are the following:
 \newline
\newline
Ark120, Circinus, ESO558-009, Fairall9, IC1481, IC2560, Mrk0079, Mrk0202,
Mrk0590, Mrk0817, NGC0193, NGC0613, NGC1097, NGC1271, NGC1275, NGC1320,
NGC1358, NGC1386, NGC1398, NGC1667, NGC1961, NGC2179, NGC2273, NGC2748,
NGC3031, NGC3079, NGC3227, NGC3368, NGC3393, NGC3627, NGC3706, NGC3783,
NGC3953, NGC3992, NGC4051, NGC4151, NGC4253, NGC4258, NGC4303, NGC4388,
NGC4429, NGC4501, NGC4507, NGC4548, NGC4593, NGC4594, NGC4698, NGC4748,
NGC4945, NGC5005, NGC5055, NGC5248, NGC5495, NGC5548, NGC5695, NGC5728,
NGC5765B, NGC6240(S), NGC6323, NGC6500, NGC6814, NGC7331, NGC7582, NGC7682,
UGC3789, UGC6093. \newline
\newline

\section{Prediction of Black Hole mass\\}

Following \citet{Graham2011} we consider the linear equation
\begin{equation}
y=\left( b\pm \delta
b\right) \left( x\pm \delta x\right) +\left( a\pm \delta a\right)
\end{equation}
with an
error on $y$ equal to

\begin{equation}
\delta y=\sqrt{\left( \frac{dy}{db}\right) ^{2}\left( \delta b\right)
^{2}+\left( \frac{dy}{da}\right) ^{2}\left( \delta a\right) ^{2}+\left(
\frac{dy}{dx}\right) ^{2}\left( \delta x\right) ^{2}}= \sqrt{x^{2}\left(
\delta b\right) ^{2}+\left( \delta a\right) ^{2}+b^{2}\left( \delta x\right)
^{2}}
\end{equation}

If we have an intrinsic scatter $\epsilon $ in the $y-$direction,
uncertainty on $y$ is
\begin{equation}
\delta y=\sqrt{x^{2}\left( \delta b\right) ^{2}+\left( \delta a\right)
^{2}+b^{2}\left( \delta x\right) ^{2}+\epsilon ^{2}}
\end{equation}
Considering the  $M_{\bullet}-\sigma $ relation for our sample of 113 early-type galaxies,

\begin{equation}
\log (M_{\bullet})=\left(-3.080\pm 0.574\right) +\left( 4.996\pm 0.248\right) \log
\left(\sigma\right)
\end{equation}
with $\epsilon = 0.365$,  we have that $x=\log
\left( \sigma \right) $,  and therefore
 considering the data of $NGC1660$ in the data base of \citet{vandenBosch2016}
and following \citet{Graham2011} we can calculate
\begin{equation}
(\delta \log (M_{\bullet}))^{2} =(6.1009) (0.0615) +0.32947+24.96(0.0004)+0.133=0.84745
\end{equation}
hence
\begin{equation}
\delta \log (M_{\bullet})=0.92057
\end{equation}
and

\begin{equation}
\log (M_{\bullet})=9.26\pm 0.92
\end{equation}

On the contrary using our best fit line obtained with MPFITEXY (Table 1)
\begin{equation}
\log \left( \frac{M_{\bullet}}{M_{\odot}}\right)=\left(0.93\pm 0.468\right) + \left( 1.0117\pm 0.0628\right) \log
\left( \frac{R\sigma ^{3}}{GcM_{\odot}}\right)
\end{equation}
with
\begin{equation}
\epsilon =0.448
\end{equation}
\begin{equation}
\log (M_{\bullet})=0.93+\left(8.378\right) \left(1.0117\right)=9.406
\end{equation}
while
\begin{equation}
(\delta \log (M_{\bullet}))^{2}=(8.378)^{2} (0.0628)^{2}+(0.468)^{2}+(1.0117)^{2}(0.11)^{2}+(0.448)^{2}=0.7088
\end{equation}
hence
\begin{equation}
\delta \log (M_{\bullet})=0.8419
\end{equation}
and
\begin{equation}
\log (M_{\bullet})=9.41\pm 0.84
\end{equation}

\end{document}